\documentclass[11pt]{article}

\parskip=\baselineskip

\usepackage{amsmath, amsthm, amssymb,slashed,mathtools,tabu}

\usepackage[usenames, dvipsnames]{color}
\usepackage[svgnames]{xcolor}
\usepackage[colorlinks,citecolor=RoyalBlue, urlcolor=RoyalBlue, linkcolor=RoyalBlue ]{hyperref} 

\usepackage{upgreek}
\usepackage{bbm}

\newenvironment{myfont}[2][]{\csname#2\endcsname[#1]}{}

\usepackage{slashed}
\usepackage[makeroom]{cancel}
\usepackage[normalem]{ulem}
\usepackage{soul}
\newcommand{\stkout}[1]{\ifmmode\text{\sout{\ensuremath{#1}}}\else\sout{#1}\fi}

\usepackage{sseq}
\usepackage[all,cmtip]{xy}
\usepackage{tikz-cd}
\usepackage{tikz}
\usetikzlibrary{matrix}
\usetikzlibrary{decorations.markings}
\usetikzlibrary{tikzmark,decorations.pathreplacing,positioning}
\usepackage{amsfonts}

\newcommand{\bea}{\begin{eqnarray}}
\newcommand{\eea}{\end{eqnarray}}
\def\be{\begin{equation}}
\def\ee{\end{equation}}

\newcommand{\ii}{\hspace{1pt}\mathrm{i}\hspace{1pt}}

\definecolor{red}{rgb}{1,0,0}
\definecolor{blue}{rgb}{0,0,1}
\definecolor{dblue}{rgb}{0,0,0.4}
\definecolor{green}{rgb}{0,1,0}
\definecolor{black}{rgb}{0,0,0}
\definecolor{white}{rgb}{1,1,1}

\definecolor{brn}{rgb}{.8,.4,.0}
\definecolor{redo}{rgb}{1,.5,.0}
\definecolor{ddgrn}{rgb}{0,0.4,0}
\definecolor{dgrn}{rgb}{0,0.55,0}
\definecolor{dbl}{rgb}{0,0,0.5}

\usepackage[bbgreekl]{mathbbol}
\usepackage{amscd}

\newcommand{\Z}{\mathbb{Z}}

\newcommand{\bpm}{\begin{pmatrix}}
\newcommand{\epm}{\end{pmatrix}}
\newcommand{\bmm}{\begin{matrix}}
\newcommand{\emm}{\end{matrix}}

\def\CN{{\cal N}}

\def\Z{{\mathbb{Z}}}

\def \Z{\mathbb{Z}}

\newcommand {\emptycomment}[1]{}

\newcommand{\U}{{\rm U}}
\newcommand{\SU}{{\rm SU}}

\newcommand{\CKM}{{\rm CKM}}
\newcommand{\UV}{{\rm UV}}
\newcommand{\IR}{{\rm IR}}

\newcommand{\Fig}[1]{Fig.~\ref{#1}} 
\newcommand{\Table}[1]{Table \ref{#1}}

\newcommand{\SM}{{\rm SM}}

\usepackage{comment}

\begin{document}

\title{\vbox{
\baselineskip 14pt
\hfill \hbox{\normalsize KEK-TH-2452
}} \vskip 1cm
\textbf{\Large Flavor Hierarchy from Smooth Confinement}
\vskip 0.5cm
}

\author{
{\bf Yuta Hamada}\thanks{E-mail: \tt yhamada@post.kek.jp}\, and {\bf Juven Wang}\thanks{E-mail: \tt jw@cmsa.fas.harvard.edu}
\vspace{5mm}
\\
$^\star$\normalsize{\it Theory Center, IPNS, High Energy Accelerator Research Organization (KEK),} \\
\normalsize{\it 1-1 Oho, Tsukuba, Ibaraki 305-0801, Japan} \\
$^\star$\normalsize{\it Graduate University for Advanced Studies (Sokendai), 1-1 Oho, Tsukuba, }
\\
\normalsize{\it Ibaraki 305-0801, Japan}\vspace{1mm} \\
$^\star$\normalsize{\it Department of Physics, Harvard University, Cambridge, MA 02138 USA}
\vspace{1mm} \\
$^\dagger$\normalsize\emph{Center of Mathematical Sciences and Applications, Harvard University, MA 02138, USA}  \vspace{1mm}
}

\date{}

\maketitle
\vspace*{1cm}

\begin{abstract} 

We present a model to explain the Standard Model flavor hierarchy.
Our model is based on explicit smooth confinement 
(namely confinement without chiral symmetry breaking in a supersymmetric gauge theory) at an intermediate energy scale, before 
the electroweak symmetry breaking by the Higgs condensation at 
lower energy. In our context, the smooth confinement preserves  
the SU(3) and the chiral SU(2)$_L\times$ U(1)$_Y$ symmetry in a supersymmetric
Standard Model, while this internal symmetry becomes dynamically gauged in the end. 
In contrast to Razamat-Tong's symmetric mass generation model also
preserving the $G_{\rm SM} \equiv$ SU(3)$_C\times$ SU(2)$_L\times$ U(1)$_Y$ internal symmetry, our model 
introduces different matter contents with a different kind 
of superpotential deformation irrelevant at UV, 
which further induces Yukawa-Higgs terms marginal at IR,
breaking the $G_{\rm SM}$  
down to SU(3)$_C$ and the
electromagnetic U(1)$_{\rm EM}$ only when Higgs condenses.
In our model, 
the IR fermions in the first and second families are 
composite of UV fields, while the third family 
elementary fermions match between UV and IR.
The smallness of the first and second family fermion masses is explained by the exponential hierarchy between the cutoff scale and the smooth confinement scale via dimensional transmutation.
As a result, our UV Lagrangian only contains the natural parameters close to the order one.

\end{abstract}



\newpage  

\section{Introduction}
One of the mysteries of the Standard Model (SM) is the flavor hierarchical structure.
The observed lightest charged fermion in our world is the electron, and its mass is 511~keV~\cite{ParticleDataGroup:2020ssz}.
On the other hand, the heaviest one is the top quark, and its mass is about 172 GeV 
\cite{deBlas:2022hdk}.
Both fermions acquire masses from the Yukawa coupling with the condensation of the Higgs field.
Therefore, the huge difference in the fermion masses indicates that the Yukawa coupling has a hierarchical structure.
More concretely, the ratio between the electron and top Yukawa coupling is $3\times 10^5$. Why is there such a large flavor hierarchy? 

One of the known solutions was provided by Froggatt and Nielsen in Ref.~\cite{Froggatt:1978nt}.
They introduced a flavor-dependent $\U(1)$ symmetry and a scalar field 
called flavon charged under it. 
The small parameter is introduced if the vev of the Higgs is smaller than the cutoff scale.

 {In this work, we present a new solution, inspired by 
the dynamics of the nonperturbative 
strong coupling of gauge theory,
known as the smooth confinement (or the s-confinement \cite{Seiberg:1994bz,Seiberg:1994pq},
which means confinement without chiral symmetry breaking in the supersymmetry context).
The idea is that we hypothesize some matter fields of the (supersymmetric) SM
at IR can be obtained as the composite fields of UV under s-confinement.
Moreover, the SM gauge group 
$G_{\rm SM} \equiv \SU(3)_C\times \SU(2)_L\times \U(1)_Y$ 
(including the chiral $\SU(2)_L\times \U(1)_Y$) 
is preserved  \emph{not} spontaneously broken by the s-confinement dynamics.}

  {Recently, the s-confinement is also applied to another problem:
the symmetric mass generation (see \cite{WangYou2204.14271} for an overview)
of the SM in four dimensions \cite{RazamatTong2009.05037,Tong:2021phe}.
It has been previously stressed that there is no obstruction to give masses and energy gaps
to all of the SM fermions without breaking the 
$G_{\rm SM}$ or Grand Unified Theory's internal symmetry 
as long as there are no anomalies among these internal symmetries \cite{WangWen2018cai1809.11171}
(see the systematic checks on the local and global anomaly cancellations for the SM in \cite{GarciaEtxebarriaMontero2018ajm1808.00009, DavighiGripaiosLohitsiri2019rcd1910.11277, WW2019fxh1910.14668} via cobordism).
Ref.~\cite{RazamatTong2009.05037,Tong:2021phe} provides an explicit model realizing this idea, in the context of the supersymmetric or non-supersymmetric SM.
The key idea in \cite{RazamatTong2009.05037,Tong:2021phe} is that the \emph{dangerously irrelevant} interaction term in the UV becomes the $G_{\rm SM}$-preserving
\emph{relevant} mass term to achieve the symmetric mass generation in the IR.}

\begin{figure}[!h]
\begin{center}
(a) \includegraphics[height=4cm]{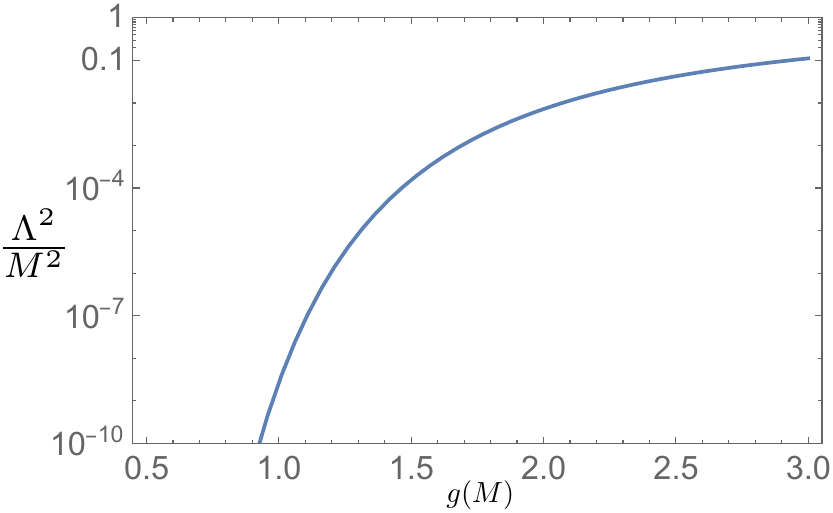} 
\quad
(b) \includegraphics[height=5.5cm]{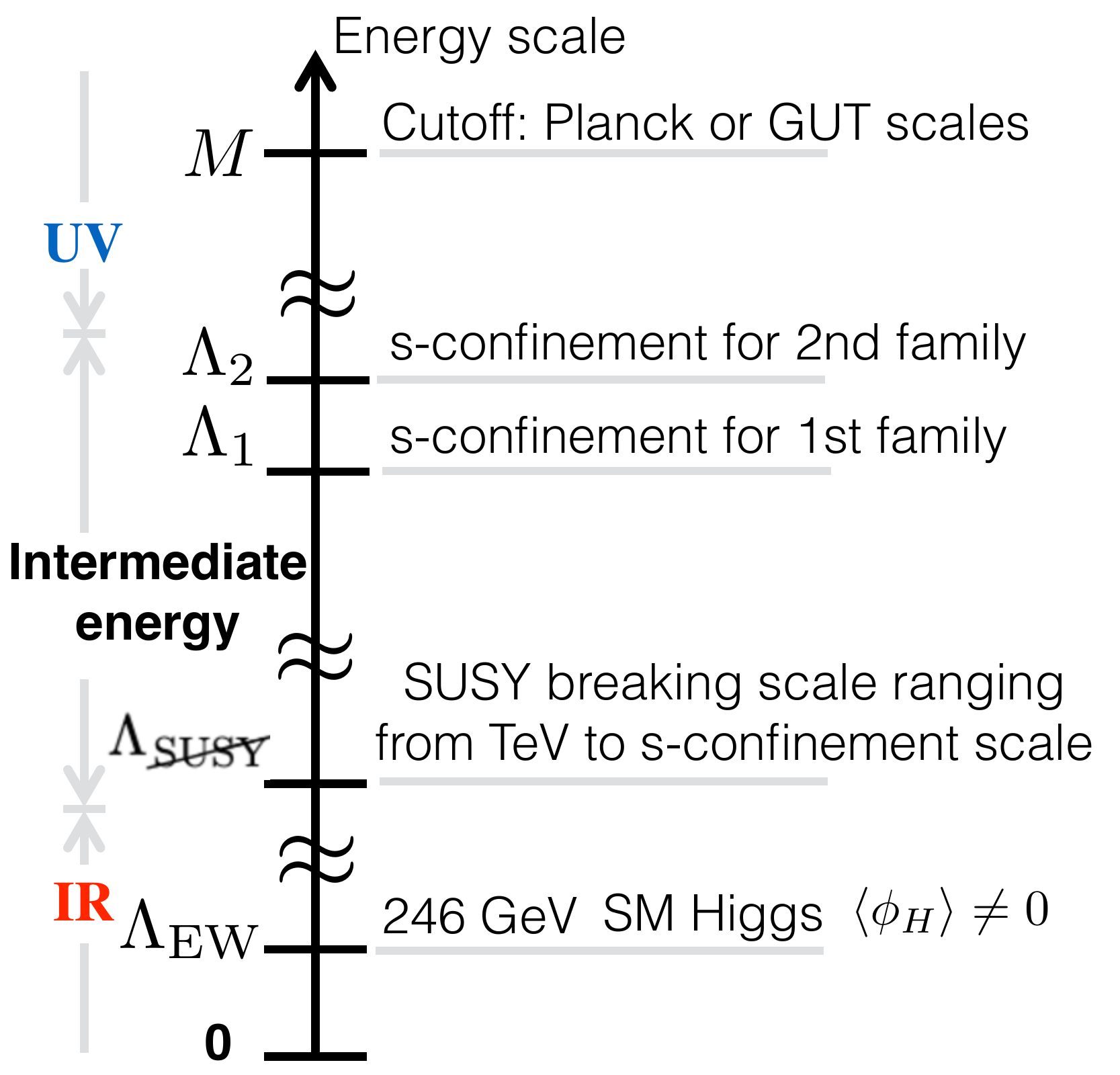}
\end{center}
\caption{(a) Ratio between the confinement and cutoff scales as a function of the gauge coupling $g(M)$ at the cutoff scale $M$. {This figure reveals that the
tiny change 
of $g(M)$ coupling
can correspond to a huge exponential hierarchy
between the confinement and the cutoff scales $\frac{\Lambda}{M}$.}
(b) Physics at different energy scales: The UV Lagrangian is valid somewhere below the cutoff $M$.
The s-confinement 
happens at intermediate energy $\Lambda_1, \Lambda_2$ while the $G_{\rm SM}$ is still preserved, 
e.g., $\frac{\Lambda_1}{M}, \frac{\Lambda_2}{M}$ ranges from $10^{-3}$ to $10^{-1}$ for phenomenological fitting.
Below the supersymmetry breaking scale $\Lambda_{\cancel{\rm SUSY}}$, 
{the SM Higgs $\phi_H$ (given by a linear combination of $H_u$ and $H_d$)
condenses
breaking $G_{\rm SM}$ to U(1)$_{\rm EM}$ below 
$\Lambda_{\rm EW}$.}
}
\label{fig:potential}
\end{figure}

  {Although both 
Razamat-Tong \cite{RazamatTong2009.05037} and our models use the s-confinement dynamics at an intermediate energy scale,
the two models (including UV Lagrangian, the deformations, 
and the IR dynamics) are rather different.
In comparison,
Ref.~\cite{RazamatTong2009.05037} considered  
enlarging from the SM's $15 N_f$ or $16 N_f$ Weyl fermions to the $27 N_f$ Weyl fermions 
that can be symmetrically gapped out by preserving $G_{\rm SM}$
via a symmetric mass generation deformation, with the family number $N_f$.
Instead, we consider 
a different UV model with 63 Weyl fermions 
in total (see Table~\ref{tab:UV_matter}, including the $15 N_f$ Weyl fermions in $N_f=3$ families)
with a different deformation. In our scenario, 
the standard Yukawa coupling at IR corresponds to the higher-scaling-dimensional operator at UV. The $G_{\rm SM}$-preserving 
\emph{dangerously irrelevant} higher-scaling-dimensional operator at UV 
becomes the \emph{marginal} Yukawa coupling at IR.
Only when the Higgs field condenses to get a vev at low energy,
then the IR dynamics further spontaneously breaks $G_{\rm SM}$ down to
the electromagnetic U(1)$_{\rm EM}$.
This naturally explains the smallness of the first and second family Yukawa couplings, where the smallness is a consequence by the ratio between the s-confinement and cutoff scales (see \Fig{fig:potential} (a)).
On the other hand, the third family fields are treated as elementary, which is consistent with $\mathcal{O}(1)$ top Yukawa coupling.
We present an explicit model to demonstrate the idea.
We show that the flavor hierarchy gets milder in our UV model, 
  the coupling of the first and second family (i.e., superpotential coefficients), and the third family Yukawa coupling
become closer to $\mathcal{O}(1)$ at high energy. Namely, 
our UV Lagrangian only contains the natural parameters
close to the order one. The physics at different energy scales from UV to intermediate to IR are shown in \Fig{fig:potential} (b).
}

\begin{table}[!t]
\centering
\begin{tabular}{ccccccccc}
 Field&  $\SU(3)_C$&  $\SU(2)_L$&  $\U(1)_Y$& $\SU(2)_1^\prime$ & $\SU(2)_2^\prime$ & $\U(1)_{B-L}$ & $\mathbb{Z}_{4,B-L}$ \\ \hline
 $L^{\prime}_{I=1}$&  $\mathbf{1}$ &  $\mathbf{2}$ & {$1/2$} & $\mathbf{2}$ & $\mathbf{1}$ & $-3$ & $1$ \\
 $L^{\prime}_{I=2}$&  $\mathbf{1}$ &  $\mathbf{2}$ &  {$1/2$} & $\mathbf{1}$ & $\mathbf{2}$ & $-3$ & $1$ \\
 $L_{I=1,2,3}$&  $\mathbf{1}$ &  $\mathbf{2}$ &  {$-1/2$} & $\mathbf{1}$ & $\mathbf{1}$ & $6$ & $2$ \\
 $D^{\prime}_{I=1}$& $\mathbf{3}$ & $\mathbf{1}$ &  {$-1/3$} & $\mathbf{2}$ & $\mathbf{1}$ & $1$ & $1$ \\
 $D^{\prime}_{I=2}$& $\mathbf{3}$ & $\mathbf{1}$ &  {$-1/3$} & $\mathbf{1}$ & $\mathbf{2}$ & $1$ & $1$ \\
 $D_{I=1,2,3}$& $\mathbf{\bar{3}}$ & $\mathbf{1}$ &  {$1/3$} & $\mathbf{1}$ & $\mathbf{1}$ & $2$ & $2$ \\
 $S_{I=1}^\prime$&  $\mathbf{1}$ & $\mathbf{1}$ &  {$0$}  & $\mathbf{2}$ & $\mathbf{1}$ & $3$ & $3$ \\
 $S_{I=2}^\prime$&  $\mathbf{1}$ & $\mathbf{1}$ &  {$0$}  & $\mathbf{1}$ & $\mathbf{2}$ & $3$ & $3$ \\
 $E_3$ & $\mathbf{1}$ & $\mathbf{1}$ &  {$1$} & $\mathbf{1}$ & $\mathbf{1}$ & $-6$ & $2$ \\
 $U_3$ & $\mathbf{\bar{3}}$ & $\mathbf{1}$ &  {$-2/3$} & $\mathbf{1}$ & $\mathbf{1}$ & $2$ & $2$ \\
 $Q_3$ & $\mathbf{3}$ & $\mathbf{2}$ &  {$1/6$} & $\mathbf{1}$ & $\mathbf{1}$ & $-2$ & $2$ \\ 
 $H_u$ & $\mathbf{1}$ & $\mathbf{2}$ &  {$1/2$} & $\mathbf{1}$ & $\mathbf{1}$ & $0$ & $0$ \\
 $H_d$ & $\mathbf{1}$ & $\mathbf{2}$ &  {$-1/2$} & $\mathbf{1}$ & $\mathbf{1}$ & $0$ & $0$\\
 $\tilde{D}_{I=1,2}$& $\mathbf{\bar{3}}$ & $\mathbf{1}$ &  {$1/3$} & $\mathbf{1}$ & $\mathbf{1}$ & $-4$ & $0$   \\
 $\tilde{L}_{I=1,2}$& $\mathbf{1}$ & $\mathbf{2}$ &   {$-1/2$} & $\mathbf{1}$ & $\mathbf{1}$ & $0$ & $0$   \\
\end{tabular}
\caption{The UV matter contents of superfields of our
$\CN=1$ minimal supersymmetric model in the left-handed basis.
$\U(1)_{B-L}$ is the anomaly free global symmetry, and the last column is $\mathbb{Z}_{4,B-L}$ subgroup of $\U(1)_{B-L}$ symmetry.
There are 22 Weyl fermions from the chiral multiplets for each of the first and second family ($I=1,2$). There are 15 Weyl fermions from the chiral multiplets for the third family ($I=3$). Each of the two Higgs multiplets introduces 2  Weyl fermions. So the model contains 63 Weyl fermions in total.
If we include additional 3 families of right-handed neutrinos neutral to
$G_{\rm SM}$, there will be 66 Weyl fermions in total.}
\label{tab:UV_matter}
\end{table}

Let us also compare 
Froggatt-Nielsen model \cite{Froggatt:1978nt} with
our model. 
Froggatt-Nielsen introduced the flavon field, which suffers from the hierarchy problem associated with the quadratic divergence to the mass squared. Moreover, it is not clear why the vacuum expectation value of the flavon is 
smaller than the cutoff scale.
One of the advantages of our model is that the new hierarchy problem does not occur, since we use s-confinement rather than the Higgs mechanism. The confinement scale appears as the result of the dimensional transmutation, and is naturally smaller than the cutoff scale.

The idea of applying the composite fermions \cite{DimopoulosRabySusskindDRS1980} 
and smooth confinement \cite{Seiberg:1994bz,Seiberg:1994pq} 
to fermion mass hierarchy dates back to Strassler \cite{Strassler19959510342}, 
Nelson and others 
\cite{Nelson19969607362}. 
There were also other earlier supersymmetric composite models based on technicolor \cite{FarhiSusskindTechnicolor1981} with similar goals. 
However, there are significant differences between our model and other previous models: 
1. We only require the $\SU(3) \times \SU(2) \times \U(1)$ SM instead of the Grand Unified Theory (GUT) group. 
2. We choose the third family fermions elementary, 
while previous works pursue different routes: such as
top-Yukawa-Higgs coupling is dynamically generated, 
top quark and bottom quark are composite \cite{Nelson19969607362}. 
3. We choose the large smooth confinement $\Lambda \simeq 10^{-3}M$ to $10^{-1} M$ close to the cutoff scale $M$,
while others choose the much smaller TeV confinement scale \cite{Nelson19969607362}.
4. The smooth confinement scale is also  
the symmetric mass generation scale in our theory, 
which naturally demands that the coupling strengths of
the original SM Lagrangian $g_{\rm SM}$
and the deformation interaction terms $g_{\rm int}$
have an order 1 ratio at our UV theory. 
This ${g_{\rm int}}/{g_{\rm SM}}\sim \mathcal{O}(1)$ 
(or more precisely the energy ratio of the action ${E_{\rm int}}/{E_{\rm SM}}\sim \mathcal{O}(1)$)
is due to the fact that the symmetric energy gap is generated nonperturbatively,
so we should \emph{not} rely on a perturbative renormalization group analysis around fixed points at 
either limit ${g_{\rm int}}/{g_{\rm SM}}\ll 1$ or ${g_{\rm int}}/{g_{\rm SM}}\gg 1$.

This article is organized as follows.
The UV model is provided in Sec.~\ref{Sec:Model}.
The model exhibits the s-confinement at the intermediate energy scale, and the minimal supersymmetric SM is realized in the IR, as we show in Sec.~\ref{Sec:IR}. 
We show that the flavor hierarchy is solved by comparing with the experimental data in Sec.~\ref{Sec:Comparison}.
The summary and discussions are in Sec.~\ref{Sec:Summary}.

\section{Model in the UV}\label{Sec:Model}

Inspired by the symmetric mass generation in four dimensions \cite{RazamatTong2009.05037}, 
we propose a new model which reduces to the supersymmetric version of the SM in the IR. 

The UV matter contents of our model is listed in Table~\ref{tab:UV_matter}.
Regarding the first and second families, we introduce the superfields $(L^{\prime}_I,D^{\prime}_I,S_I^\prime)$ charged under $\SU(2)_1^\prime\times \SU(2)_2^\prime$ ($I=1,2$ is the index for the family).
There are also superfields $(D_I, \tilde{D}_I, L_I, \tilde{L}_I)$, which are neutral under $\SU(2)_1^\prime\times \SU(2)_2^\prime$.
As for the third family, the matter contents are the same as those of the $\CN=1$ Minimal Supersymmetric SM, $(Q_3, U_3, D_3, L_3, E_3)$.
On top of that, we introduce the Higgs superfields $(H_u, H_d)$. These are necessary to cancel the dynamical gauge anomaly of 
$G_{\rm SM} \equiv \SU(3)_C\times \SU(2)_L\times \U(1)_Y$.
In total, there are $63$ Weyl fermions in Table~\ref{tab:UV_matter}. 

The superpotential of the model is\footnote{To be precise, more terms are allowed since $H_d$ and $\tilde{L}_I$ have the same quantum number. Similarly, $H_u$ and $L^\prime_I S_I^\prime$ has the same quantum number. 
In generic situation, the last two terms in Eq.~\eqref{Eq:UV_superpotential} will be 3 by 3 mass matrix in the IR. We assume that one of the linear combination is relatively light, which we call $H_u$ and $H_d$. {The electroweak Higgs is obtained from a linear-combination of $H_u$ and $H_d$.}}
\begin{align}
    W_{\UV}=&y^u_{IJ}\frac{(L^\prime_I D^\prime_I)(D^\prime_JD^\prime_J)H_u}{M^2}
    +y^d_{IJ}\frac{(L^\prime_ID^\prime_I)D_J H_d}{M}
    +y^l_{IJ} \frac{L_I(L^\prime_JL^\prime_J)H_d}{M}
    \nonumber\\
    &+y_{3I}^u \frac{Q_3 (D^\prime_ID^\prime_I)H_u}{M}
    +y_{I3}^u \frac{(L^\prime_I D^\prime_I) U_3 H_u}{M}
    +y^d_{3I} Q_3 D_I H_d
    +y^d_{I3} \frac{(L^\prime_ID^\prime_I)D_3 H_d}{M}
    \nonumber\\
    &+y^l_{3I} \frac{L_3(L^\prime_IL^\prime_I)H_d}{M}
    +y^l_{I3} L_I E_3 H_d
    +y_{33}^u Q_3 U_3 H_u+ y_{33}^d Q_3 D_3 H_d 
    \nonumber\\
    &+ y_{33}^l L_3 E_3 H_d
    +\lambda^D_{I}\tilde{D}_I (D_I^\prime S_I^\prime)
    +\mu H_u H_d
    +\lambda^L_{IJ} \tilde{L}_I (L_J^\prime S_J^\prime)
\label{Eq:UV_superpotential}\end{align}
where $M$ is the cutoff scale of the model. This could be the GUT
or Planck scale.
This is the most general superpotential 
(up to four-fermion interaction terms)
which possess $\U(1)_{B-L}$ (or its subgroup $\mathbb{Z}_{4,B-L}$)
defined in Table~\ref{tab:UV_matter}.
As we will see, the $\mathbb{Z}_{4,B-L}$ subgroup of $\U(1)_{B-L}$ is identified as matter parity in the Minimal Supersymmetric SM.
If we do not impose $\U(1)_{B-L}$ or $\mathbb{Z}_{4,B-L}$, the following superpotential is allowed:
\begin{align}
    &W_{\Delta L=1,\UV}=\lambda^{IJK}\frac{L_IL_J(L^\prime_KL^\prime_K)}{M}
    +\lambda^{\prime IJK}\frac{L_I (L^\prime_JD^\prime_J)D_K}{M}
    +\mu^{\prime I}L_IH_u
    +\mu^{\prime\prime IJ}L_I (L_J^\prime S_J^\prime),
    \nonumber\\
    &W_{\Delta B=1,\UV}= \lambda^{\prime\prime IJK} \frac{(D_I^\prime D^\prime_I) D_J D_K}{M}.
\label{Eq:W_BLviolating}\end{align}
{These terms must be small in order to avoid the fast proton decay.}

\section{Effective theory in the IR}\label{Sec:IR}

The beta function of $\SU(2)_1^\prime\times \SU(2)_2^\prime$ gauge coupling is negative, and therefore the system is strongly coupled at IR, and asymptotic free at UV.
The one-loop beta functions of the model is
\begin{align}
    &\mu\frac{dg_I}{d\mu}
=-\left(\frac{11}{3}C_2(G)-\frac{1}{3}n_sT(R_s)-\frac{2}{3}  2n_f  T(R_f)\right)\frac{g_I^3}{16\pi^2}
    =-\frac{13 g_I^3}{48 \pi^2},
    &&I=1,2.
\end{align}
where $g_I$ is the gauge coupling of $\SU(2)_I^\prime$ at the energy scale $\mu$,
the quadratic Casimir is $C_2(G)=2$ of $G=\SU(2)'_I$.
The number of complex scalars $n_s=6$ and the 
number of Weyl fermions $2n_f=6$ that couple to $G$ are all from the 
6 chiral multiplets. $T(R_s)=T(R_f)=\frac{1}{2}$ for the fundamental representation.
The strongly coupled s-confinement scale $\Lambda_I$ is
\begin{align}
    \Lambda_I^2=M^2 \exp\left(-\frac{48\pi^2}{13g_I^2(M)}\right),
\end{align}
where $g_I(M)$ is the coupling at the cutoff scale $\mu=M$.

In the IR, the system exhibits the s-confinement~\cite{Seiberg:1994bz,Seiberg:1994pq}, which magic requires
$n_c+1=n_f$ where $n_c=2$ is from $\SU(2)_I'$ and $n_f=3$ implies the 6 Weyl fermions
that couple to the $\SU(2)_I'$.
{The IR fields are the composites of UV fields:}
\begin{align}
    E_I\equiv \frac{\epsilon_{ab}L^{\prime a}_I L^{\prime b}_I}{\Lambda_I},
    \;\;
    U_{kI}\equiv\frac{\epsilon_{ijk}D^{\prime i}_I D^{\prime j}_I}{\Lambda_I},
    \;\;
    Q_{bI}^i\equiv\frac{\epsilon_{ab}L^{\prime a}_ID^{\prime i}_I}{\Lambda_I},
    \;\;
    \tilde{D}_I^{\prime i}\equiv\frac{D^{\prime i}_I S^\prime_I}{\Lambda_I},
    \;\;
    \tilde{L}^\prime_I\equiv\frac{L^\prime_I S^\prime_I}{\Lambda_I},
\label{Eq:IR}\end{align}
where $a,b=1,2$ is $\SU(2)_L$ index, $i,j,k=1,2,3$ is $\SU(3)_C$ index, and $\Lambda_I$ is the confinement scale of $\SU(2)_I^\prime$.
The IR matter contents are summarized in \Table{tab:IR_matter}.

At IR low energy below the s-confinement scale, 
the superpotential~\eqref{Eq:UV_superpotential} becomes
\begin{align}
    &W_{\IR}=y^u_{IJ} \frac{\Lambda_I \Lambda_J}{M^2} Q_I U_J H_u
    +y^d_{IJ} \frac{\Lambda_I}{M} Q_I D_J H_d
    +y^l_{IJ} \frac{\Lambda_J}{M} L_I E_J H_d
    \nonumber\\
    &+y_{3I}^u \frac{\Lambda_I}{M} Q_3 U_I H_u
    +y_{I3}^u \frac{\Lambda_I}{M} Q_I U_3 H_u
    +y^d_{3I} Q_3 D_I H_d
    +y^d_{I3} \frac{\Lambda_I}{M} Q_I D_3 H_d
    \nonumber\\
    &+y^l_{3I} \frac{\Lambda_I}{M} L_3E_IH_d
    +y^l_{I3}  L_I E_3 H_d
    +y_{33}^u Q_3 U_3 H_u
    + y_{33}^d  Q_3 D_3 H_d 
    \nonumber\\
    &+ y_{33}^l L_3 E_3 H_d
    +\mu H_u H_d
    +\lambda^D_{I} \Lambda_I \tilde{D}_I \tilde{D}^\prime_I
    +\lambda^L_I \Lambda_I \tilde{L}_I \tilde{L}_I
    \nonumber\\
    &\equiv\tilde{y}^{u}_{\alpha\beta} Q_\alpha U_\beta H_u
    +\tilde{y}^{d}_{\alpha\beta} Q_\alpha D_\beta H_d
    +\tilde{y}^{l}_{\alpha\beta} L_\alpha E_\beta H_d
    +\tilde{\mu}\, H_u H_d
    +\tilde{M}^D_{I} \tilde{D}_I \tilde{D}^\prime_I
    +\tilde{M}^L_I \,\tilde{L}_I \tilde{L}^\prime_I,
\label{Eq:IR_superpotential}\end{align}
where $\alpha,\beta=1,2,3$, while $\tilde{y}^{u}, \tilde{y}^{d}$ and $\tilde{y}^{l}$ are the low-energy Yukawa couplings.

In the IR, Eq.~\eqref{Eq:W_BLviolating} becomes
\begin{align}
    &W_{\Delta L=1,{\rm IR}}=\lambda^{IJK}L_IL_JE_K
    +\lambda^{\prime IJK}L_I Q_J D_K
    +\mu^{\prime I}L_IH_u
    +\mu^{\prime\prime IJ}L_I \tilde{L}_J^\prime,
    \nonumber\\
    &W_{\Delta B=1,{\rm IR}}= \lambda^{\prime\prime IJK} U_{I} D_J D_K.
\end{align}
Again, this breaks $\U(1)_{B-L}$ and its $\mathbb{Z}_{4,B-L}$ subgroup.

\section{Comparison with experimental data}\label{Sec:Comparison}
From \cite{Antusch:2013jca}, the experimental data of the Yukawa coupling is\footnote{Here $\tan\beta$ is defined as the ratio of the VEV of two Higgs at the electroweak symmetry breaking vacuum, $\tan\beta \equiv \langle H_u\rangle/\langle H_d\rangle$.}
\begin{align}
    &\tilde{y}^u_{11}\sin\beta\sim 6\times 10^{-6},
    &&\tilde{y}^u_{22}\sin\beta\sim 3\times 10^{-3},
    &&\tilde{y}^u_{33}\sin\beta\sim 0.8,
    \nonumber\\
    &\tilde{y}^d_{11}\cos\beta\sim 1\times 10^{-5},
    &&\tilde{y}^d_{22}\cos\beta \sim 3\times 10^{-4},
    &&\tilde{y}^d_{33}\cos\beta\sim 1\times 10^{-2},
    \nonumber\\
    &\tilde{y}^l_{11}\cos\beta\sim 3\times 10^{-6}
    &&\tilde{y}^l_{22}\cos\beta\sim 6\times 10^{-4}
    &&\tilde{y}^l_{33}\cos\beta\sim 1\times 10^{-2}.
\end{align}

\begin{table}[t]
\centering
\begin{tabular}{cccccccc}
 Field&  $\SU(3)_C$&  $\SU(2)_L$&  $\U(1)_Y$& $\U(1)_{B-L}$ & $\mathbb{Z}_{4, B-L}$ & $\mathbb{Z}_{2,{\bf R}}$ \\ \hline
 $E_{I=1,2,3}$&  $\mathbf{1}$ &  $\mathbf{1}$ &  {$1$} &  $-6$ & $2$ &  1 \\
 $U_{I=1,2,3}$&  $\mathbf{\bar{3}}$ &  $\mathbf{1}$ &  {$-2/3$} & $2$ & $2$ &  1 \\
 $Q_{I=1,2,3}$&  $\mathbf{3}$ &  $\mathbf{2}$ &  {$1/6$} & $-2$ & $2$ & 1 \\ 
 $L_{I=1,2,3}$&  $\mathbf{1}$ &  $\mathbf{2}$ &  {$-1/2$} &  $6$ & $2$ & 1 \\ 
 $D_{I=1,2,3}$&  $\mathbf{\bar{3}}$ &  $\mathbf{1}$ &  {$1/3$} & $2$ & $2$ & 1 \\
 $\tilde{D}_{I=1,2}$&  $\mathbf{\bar{3}}$ &  $\mathbf{1}$ &  {$1/3$} & $-4$ & $0$ & 0 \\
 $\tilde{D}^\prime_{I=1,2}$&  $\mathbf{3}$ &  $\mathbf{1}$ &  {$-1/3$} & $4$ & $0$ & 0  \\
 $H_u$ & $\mathbf{1}$ & $\mathbf{2}$ &  {$1/2$} & $0$ & $0$ & 0 \\
 $H_d$ & $\mathbf{1}$ & $\mathbf{2}$ &  {$-1/2$} & $0$ & $0$ & 0\\
 $\tilde{L}_I$ & $\mathbf{1}$ & $\mathbf{2}$ &  {$-1/2$} & $0$ & $0$  & 0 \\
 $\tilde{L}^\prime_I$ & $\mathbf{1}$ & $\mathbf{2}$ &  {$1/2$} & $0$ & $0$ & 0 \\
\end{tabular}
\caption{The IR matter contents in the left-handed basis. The IR composite fields are given in Eq.~\eqref{Eq:IR}.
All fields are singlet under $\SU(2)_1^\prime\times \SU(2)_2^\prime$.
The original $\mathbb{Z}_{4,B-L}$ symmetry  is identified as matter-parity $\mathbb{Z}_{2,{\bf R}}$ of the minimal supersymmetric standard model (MSSM)~\cite{Martin:1997ns} in the IR.
We can understand that the MSSM's chiral multiplets are composed of two fermions with
$\Z_{4,B-L}$ charge 1 from Table \ref{tab:UV_matter},
which become the IR field with
$\Z_{4,B-L}$ charge 2,
thus with $\mathbb{Z}_{2,{\bf R}}$ charge 1,
under the s confinement (via
$\SU(2)_1^\prime$ and $\SU(2)_2^\prime$ dynamical gauge field confinement).
Here we have a normal subgroup embedding
$\mathbb{Z}_{2,{\bf R}} \subset \Z_{4,B-L}
\subset  \U(1)_{B-L}$.
}
\label{tab:IR_matter}
\end{table}

The singular value decompositions of Yukawa coupling
$\tilde{y}^u$ and $\tilde{y}^d$ are
\begin{align}
    &\tilde{y}^u=U^u\text{diag}(Y_u, Y_c, Y_t)V^u,
    &&\tilde{y}^d=U^d\text{diag}(Y_d, Y_s, Y_b)V^d,
\end{align}
where $U^{u,d}$ and $V^{u,d}$ are unitary matrices.
The CKM matrix is defined as
\begin{align}
    V_{\CKM}=(U^u)^\dagger U^d.
\end{align}
The Wolfenstein parametrization of the CKM matrix is
\begin{align}
    V_{\CKM}&=
    \begin{pmatrix}
    1-\frac{1}{2}\lambda^2 & \lambda & A\lambda^3(\rho-\ii\eta)\\
    -\lambda& 1-\frac{1}{2}\lambda^2& A\lambda^2\\
    A\lambda^3(1-\rho-\ii\eta)& -A\lambda^2& 1
    \end{pmatrix}
    +\mathcal{O}(\lambda^4),
\label{Eq:CKM_experiment}
\end{align}
where $\lambda\simeq0.23$, $A\simeq0.8$, $\rho\simeq0.14$ and $\eta\simeq0.35$.
This means that the observed value of CKM matrix is
\begin{align}
    V_{\CKM}^\mathrm{obs}\sim
    \begin{pmatrix}
    1 & 0.2 & 0.001-0.003 \ii \\
    0.2 & 1& 0.04\\
    0.01-0.003 \ii & 0.04& 1
    \end{pmatrix}.
    \label{Eq:CKM_experiment-2}
\end{align}
Similarly, from \cite{Antusch:2013jca}, the experimental value of $Y^{u,c,t}$ and $Y^{d,s,b}$ is 
\begin{align}
    &Y_u^\mathrm{obs}\sim \frac{6\times10^{-6}}{\sin\beta},
    &&Y_c^\mathrm{obs}\sim \frac{3\times10^{-3}}{\sin\beta},
    &&Y_t^\mathrm{obs}\sim \frac{0.8}{\sin\beta},
    \nonumber\\
    &Y_d^\mathrm{obs}\sim  \frac{1\times 10^{-5}}{\cos\beta},
    &&Y_s^\mathrm{obs}\sim \frac{3\times10^{-4}}{\cos\beta},
    &&Y_b^\mathrm{obs}\sim \frac{1\times10^{-2}}{\cos\beta},
    \nonumber\\
    &Y_e^\mathrm{obs}\sim \frac{3\times10^{-6}}{\cos\beta},
    &&Y_\mu^\mathrm{obs}\sim \frac{6\times10^{-4}}{\cos\beta},
    &&Y_\tau^\mathrm{obs}\sim \frac{6\times10^{-2}}{\cos\beta}.
\label{Eq:Y_obs}\end{align}

Let us estimate the Yukawa couplings in our model.
From \eqref{Eq:IR_superpotential}, the order of the up and down quark Yukawa couplings are
\begin{align}
    &\tilde{y}^u\sim
    \begin{pmatrix}
    \dfrac{\Lambda_1^2}{M^2}
    & \dfrac{\Lambda_1\Lambda_2}{M^2}
    &\dfrac{\Lambda_1}{M}\\[0.5em]
    \dfrac{\Lambda_1\Lambda_2}{M^2}
    &\dfrac{\Lambda_2^2}{M^2}
    &\dfrac{\Lambda_2}{M}\\[0.5em]
    \dfrac{\Lambda_1}{M}
    &\dfrac{\Lambda_2}{M}
    &1
    \end{pmatrix},
    &&\tilde{y}^d\sim
    \begin{pmatrix}
    \dfrac{\Lambda_1}{M}
    &\dfrac{\Lambda_1}{M}
    &\dfrac{\Lambda_1}{M}\\[0.5em]
    \dfrac{\Lambda_2}{M}
    &\dfrac{\Lambda_2}{M}
    &\dfrac{\Lambda_2}{M}\\[0.5em]
    1
    &1
    &1
    \end{pmatrix},
\end{align}
while the charged lepton Yukawa coupling is
\begin{align}
    \tilde{y}^l\sim
    \begin{pmatrix}
    \dfrac{\Lambda_1}{M}
    &\dfrac{\Lambda_2}{M}
    &1\\[0.5em]
    \dfrac{\Lambda_1}{M}
    &\dfrac{\Lambda_2}{M}
    &1\\[0.5em]
    \dfrac{\Lambda_1}{M}
    &\dfrac{\Lambda_2}{M}
    &1
    \end{pmatrix}.
\end{align}
By performing the singular value decomposition, we obtain
\begin{align}
    &Y_u \sim \frac{\Lambda_1^2}{M^2},
    &&Y_c\sim \frac{\Lambda_2^2}{M^2},
    &&Y_t\sim 1,
    \nonumber\\
    &Y_d \sim \frac{\Lambda_1}{M^2},
    &&Y_s\sim \frac{\Lambda_2}{M},
    &&Y_b\sim 1,
    \nonumber\\
    &Y_e\sim \frac{\Lambda_1}{M} ,
    &&Y_\mu\sim \frac{\Lambda_2}{M},
    &&Y_\tau\sim 1,
\end{align}
and
\begin{align}
    &U^u\sim
    \begin{pmatrix}
    1& \dfrac{\Lambda_1}{M}& \dfrac{\Lambda_1}{M}\\[0.7em]
    \dfrac{\Lambda_1}{M}& 1& \dfrac{\Lambda_2}{M}\\[0.7em]
    \dfrac{\Lambda_1}{M}& \dfrac{\Lambda_2}{M}& 1
    \end{pmatrix},
    &&U^d\sim
    \begin{pmatrix}
    1
    & \dfrac{\Lambda_1}{\Lambda_2}
    & \dfrac{\Lambda_1}{M}\\[0.7em]
    \dfrac{\Lambda_1}{\Lambda_2}& 1
    &\dfrac{\Lambda_2}{M} \\[0.7em]
    \dfrac{\Lambda_1}{M}
    &\dfrac{\Lambda_2}{M} 
    &1 
    \end{pmatrix}.
\end{align}
Then, the CKM matrix is {(here $\mathcal{O}(N_1,N_2)$ chooses the maximal among the $N_1$ and $N_2$)}:
\begin{align*}
    V_{\CKM}=(U^u)^\dagger U^d&
    \sim
    \begin{pmatrix}
    1
    &\mathcal{O}\left(\dfrac{\Lambda_1}{M},\dfrac{\Lambda_1}{\Lambda_2}\right)
    &\dfrac{\Lambda_1}{M}\\[0.8em]
    \mathcal{O}\left(\dfrac{\Lambda_1}{M},\dfrac{\Lambda_1}{\Lambda_2}\right)
    &1
    & \dfrac{\Lambda_2}{M}\\[0.8em]
    \dfrac{\Lambda_1}{M}
    &\mathcal{O}\left(\dfrac{\Lambda_2}{M},\dfrac{\Lambda_1^2}{M\Lambda_2}\right)
    &1
    \end{pmatrix}.
\label{Eq:our_CKM}\end{align*}

If we choose
\begin{align}
    &\frac{\Lambda_1}{M} \sim 3\times10^{-3}, 
    &&\frac{\Lambda_2}{M} \sim 3\times 10^{-2},
\end{align}
then, the singular values are
\begin{align}
    &Y_u \sim 9\times10^{-6},
    &&Y_c \sim  9\times10^{-4},
    &&Y_t \sim 1,
    \nonumber\\
    &Y_d \sim 3\times10^{-3},
    &&Y_s \sim 0.03,
    &&Y_b \sim 1,
    \nonumber\\
    &Y_e \sim 3\times10^{-3} ,
    &&Y_\mu \sim 0.03,
    &&Y_\tau \sim 1.
\label{Eq:Y_numerics}\end{align}
and the CKM matrix is
\begin{align}
    V_{\CKM} \sim
    \begin{pmatrix}
    1
    & 0.1
    & 3\times10^{-3}\\
    0.1
    &1
    & 0.03\\
    3\times10^{-3}
    &0.03
    &1
    \end{pmatrix}.
\label{Eq:CKM_numerics}\end{align}

By comparing \eqref{Eq:Y_obs} and \eqref{Eq:Y_numerics}, we observe 
\begin{align}
    &\frac{Y_u}{Y_u^\mathrm{obs}} \sim 1.5\sin\beta,
    &&\frac{Y_c}{Y_c^\mathrm{obs}} \sim 0.5\sin\beta,
    &&\frac{Y_t}{Y_t^\mathrm{obs}} \sim 1\sin\beta,
    \nonumber\\
    &\frac{Y_d}{Y_d^\mathrm{obs}} \sim  300\cos\beta,
    &&\frac{Y_s}{Y_s^\mathrm{obs}}\sim  100\cos\beta,
    &&\frac{Y_b}{Y_b^\mathrm{obs}} \sim  100\cos\beta,
    \nonumber\\
    &\frac{Y_e}{Y_e^\mathrm{obs}} \sim  10^3\cos\beta,
    &&\frac{Y_\mu}{Y_\mu^\mathrm{obs}}\sim  50\cos\beta,
    &&\frac{Y_\tau}{Y_\tau^\mathrm{obs}} \sim 20\cos\beta.
\end{align}
Similarly, from \eqref{Eq:CKM_experiment-2} and \eqref{Eq:CKM_numerics}, we get the $\mathcal{O}(1)$ ratio between our theoretical fit and
the experimental data:
\begin{align}
    \frac{|V_{\CKM,IJ}|}{|V_{\CKM,IJ}^\mathrm{obs}|}
    \sim
    \begin{pmatrix}
    1& 2 & 1\\
    2 &1& 1\\
    3&1 &1
    \end{pmatrix}.
\end{align}
We observe that the CKM matrix is nicely fitted. 

As for the fermion masses, at least the hierarchy is milder than the original SM.
If we choose smaller $\cos\beta$, the hierarchy is getting milder. For example, $\cos\beta\sim0.1$\,(so $\tan\beta=10$) corresponds to $Y_e/Y_e^\mathrm{obs}\sim 10^2$, and $\cos\beta\sim0.02$\,(so $\tan\beta=40$) corresponds to $Y_e/Y_e^\mathrm{obs}\sim 20$.\footnote{Roughly speaking, the large $\tan\beta$ corresponds to a smaller supersymmetry breaking scale. Given the bound from LHC, the value of $\tan\beta$ cannot be much bigger than $50$.}

The neutrino sector is the same as the SM.
By adding the right-handed neutrino superfield which is singlet under $G_{\SM}$, we can write down the neutrino Yukawa coupling and Majorana mass:
\begin{align}
    W_N = y^N_{IJ} L_I N_J H_u + M_{IJ}N_I N_J.
\end{align}
These terms break continuous $\U(1)_{B-L}$ symmetry, but preserves $\mathbb{Z}_{4,B-L}$ symmetry by assigning charge $2$ to $N_I$. Adding 3 right-handed neutrinos would change the model from 63 to 66 Weyl fermions.

After the electroweak symmetry breaking, the neutrino gets the Dirac mass $(M_D)_{IJ}=y^N_{IJ}\langle H_u\rangle$. By integrating out right handed neutrino, the light neutrino mass matrix is
\begin{align}
    (m^\nu)_{IJ}=(m_D)_{IK}(M^{-1})_{KL}(m_D)_{LJ}.
\end{align}
This is diagonalized as
\begin{align}
    (m^\nu)_{IJ} &= U^\nu \mathrm{diag}(m^{\nu_1},m^{\nu_2},m^{\nu_3}) (U^\nu)^T.\\
    \tilde{y}^e &=U^l\text{diag}(Y_e, Y_\mu, Y_\tau)V^{l},
\end{align}
{Then, the PMNS matrix is defined as
\begin{align}
    U_{\rm PMNS}=U^\nu U^{l\dagger}
    ={\begin{pmatrix}
    c_{12}c_{13} & s_{12}c_{13} & s_{13} e^{-\ii \delta_{CP}}\\
    -s_{12}c_{23}-c_{12}s_{23}s_{13}e^{\ii \delta_{CP}}& c_{12}c_{23}-s_{12}s_{23}s_{13}e^{\ii \delta_{CP}} & s_{23}c_{13}\\
    s_{12}s_{23}-c_{12}c_{23}s_{13}e^{\ii \delta_{CP}}
    &-c_{12}s_{23}-s_{12}c_{23}s_{13}e^{\ii \delta_{CP}}& c_{23}c_{13}
    \end{pmatrix}},
\end{align}
where $s_{IJ}=\sin\theta_{IJ}, c_{IJ}=\cos\theta_{IJ}$ are the mixing angles, and $\delta_{CP}$ is the CP phase.
The matrix $V^\ell$ is constrained by the charge assignment in our model, whereas the matrix $U^\ell$ is arbitrary. Therefore, the experimental data is fitted by choosing $U^\nu$ appropriately.} Beware that our model has not yet explained the smallness of the mixing angle $\theta_{13}$, which may require a mechanism other than the one we propose.

\section{Summary}\label{Sec:Summary}
We have proposed a new mechanism involving the s-confinement 
to explain the flavor hierarchy in the Standard Model.
The s-confinement dynamics is used to drive the \emph{dangerously irrelevant} higher-scaling-dimensional
operators at UV to the 
\emph{marginal} Yukawa-Higgs deformation at IR in our model.  
(In contrast, the s-confinement is used differently in the symmetric mass generation deformation in \cite{RazamatTong2009.05037,Tong:2021phe}.)
The smallness of the Yukawa coupling in the flavor hierarchy problem
is explained as the ratio of the s-confinement and the cutoff scales.
We propose an explicit supersymmetric model to resolve the flavor hierarchy 
with the $\mathcal{O}(1)$ coefficients at UV.
{The UV Lagrangian is more consistent with 
Dirac or 't Hooft criteria on the naturalness \cite{tHooft1979Naturalness}.}
In fact, when the 
SMG happens, the relative strengths between the Standard Model action $S_{\rm SM}$ 
and the SMG interaction action $g S_{\rm SMG, int}$ together
$
S_{\rm SM} + g\, S_{\rm SMG, int}
$
have the dimensionless
coupling ratio $g \simeq \mathcal{O}(1)$ of the naturalness order 1 
(Also $|\frac{E_{\rm SMG, int}}{E_{\rm SM}}|\sim \mathcal{O}(1)$ for the energy ratios) \cite{Wang2013ytaJW1307.7480,Zeng2202.12355:2022grc}.
So in fact the SMG types of processes demand Naturalness!

{Some comments and promising future directions follow.

1. It will be interesting to investigate both the supersymmetric and non-supersymmetric scenarios of the models similar to ours for
the flavor hierarchy resolution. 

2. One can also consider the symmetric mass generation 
deformation of our 63 or 66 Weyl fermion model.
One can also enlarge our model to include also the $27 N_f$ Weyl fermion model of \cite{RazamatTong2009.05037}, 
so to introduce both the symmetric mass generation and Yukawa-Higgs deformations under the umbrella of a parent theory. 
One can study the phase transitions between different phases of the parent theory \cite{WangWanYou2112.14765}.

3. The symmetric mass generation in \cite{RazamatTong2009.05037,Tong:2021phe} still allows a 
\emph{mean-field} fermion bilinear mass term description at IR, 
although this mean-field mass does not break $G_{\SM}$ 
because that the $G_{\SM}$ is redefined when 
the matter contents are enlarged by adding more fields from UV.
However, a more intrinsic symmetric mass generation requires 
a \emph{non-mean-field} mass deformation, purely driven by the
multi-field interactions or the disorder mass field configurations \cite{WangYou2204.14271, Wang2207.14813, StrongCPtoappear} --- this alternative 
\emph{beyond-mean-field} deformation
deserves future study \cite{StrongCPtoappear}. 

4. The large $\tan \beta$ mildens the flavor hierarchy, but the
$\tan \beta$ cannot be too large otherwise it lowers supersymmetry breaking scale that violates the experimental bound. It will be desirable to sharpen the rough constraint~\cite{Degrassi:2012ry} $\tan \beta < 50$ here to find the future phenomenology evidence. 

5. The $s$-confinement scenario has the asymptotic freedom at the deep UV $\gg \Lambda$,  while the higher-scaling-dimensional operator and multi-field interaction are nonrenormalizable becoming nonperturbatively strong also at UV approaching to the cutoff scale $M$. Closer to the cutoff $M$ challenges the validity of the effective field theory, it is worthwhile to find, 
in addition to the discrete lattice formulation, 
any alternative continuum UV completion of our model.}

6. As a condensed matter application, the (iso)spin and the electrically charged degrees of freedom can be separated by a larger energy gap. It will be interesting to know if the smooth confinement analogy works or not to explain the (iso)spin-charge separation hierarchy.

\section*{Acknowledgment}
The work of YH was supported by Japan Society for the Promotion of Science (JSPS) Overseas Research Fellowships.
At the final stages of the work, YH was supported by MEXT Leading Initiative for Excellent Young Researchers Grant Number JPMXS0320210099. The work of JW is supported by Harvard University CMSA. JW thanks the participants of Generalized Global Symmetries workshop (September 19-23, 2022) at Simons Center for Geometry and Physics for the inspiring program. After the manuscript submission, we are grateful to receive helpful comments from
Nathan Seiberg and Matthew Strassler.

\bibliographystyle{TitleAndArxiv}
\bibliography{BSM-CP.bib}

\end{document}